\documentclass[cameraready]{Interspeech}

\title{Whisper Hallucination Detection and Mitigation via Hidden Representation Steering and Sparse AutoEncoders}

\author[affiliation={1,2}]{Georgii}{Aparin}
\author[affiliation={1,3}]{Vadim}{Popov}
\author[affiliation={1,3}]{Tasnima}{Sadekova}
\author[affiliation={1}]{Assel}{Yermekova}

\address{
    $^1$ AI Foundation and Algorithm Lab \\
    $^2$ National University of Science and Technology MISIS \\
    $^3$ National Research University Higher School of Economics
}

\email{aparingm@gmail.com}

\keywords{Automatic Speech Recognition, Hallucinations, Whisper, Sparse AutoEncoder, Activation Steering}

\usepackage{comment}
\usepackage{multirow}
\usepackage{hyperref}
\usepackage{url}

\begin{document}

\maketitle

\begin{abstract}
Whisper, a widely adopted ASR model, is known to suffer from hallucinations — coherent transcriptions generated for non-speech audio entirely disconnected from the input. We investigate whether hallucinations can be detected and mitigated through Whisper's internal representations. We extract audio encoder activations and evaluate two representation spaces: raw Whisper activations and Sparse AutoEncoder (SAE) latents. We show that both spaces encode linearly separable hallucination-related information, with discriminative power concentrated in a sparse feature subset and increasing toward deeper encoder layers. We propose two steering strategies: activation-space steering and SAE latent-space steering. SAE-based steering reduces hallucination rate from 72.63\% to 14.11\% for Whisper small and from 86.88\% to 27.33\% for Whisper large-v3 on the full non-speech test set, with small WER degradation on speech data, approaching the performance of fine-tuning-based methods.
\end{abstract}

\section{Introduction}

Automatic Speech Recognition (ASR) is a speech processing task with a long history, evolving from classic machine learning algorithms such as hidden Markov models (HMMs) \cite{hmm1, hmm2} and finite-state transducers \cite{wfst} to more accurate hybrid approaches combining HMMs with multi-layer perceptrons \cite{hmm-mlp1, hmm-mlp2}. 
Currently state-of-the-art ASR algorithms are solely based on more sophisticated artificial neural networks trained on large speech corpora \cite{whisper}. Moreover, now ASR models are used not only to solve this particular task in dedicated applications on devices \cite{mobile_asr}, but also as an integral part of various speech processing pipelines \cite{emilia, kimi_audio} composed of multiple modules and used to ensure high quality of data provided for training speech LLMs \cite{stepaudio2, qwen3omni}, the new frontier in speech communication between humans and top-notch chat-bots. As for dedicated ASR applications, specific scenarios such as streaming \cite{mobile_asr, long_form_asr, giga_am} and long-form ASR \cite{long_form_asr, vibevoice_asr} are now considered more and more frequently.

However, despite their superior performance compared to algorithms based on classic machine learning, neural-based ASR has been recently shown to suffer from the so-called \textit{hallucinations} similar to NLP models \cite{nlg_hallu,cureus}. Although there is no a common and well-agreed definition of what a hallucination of an ASR model is, we stick to a reasonable definition given in \cite{identifying_errors} suggesting that a hallucination is a ``fluent and coherent outputs of neural models entirely disconnected from the input''. Such a definition allows for further classification of ASR hallucinations, but in this paper we consider their most popular type \cite{calm_whisper, investigation_hallucinations}, namely assigning coherent transcripts to non-speech segments.

There are various possible ways of overcoming the mentioned drawback of ASR models. They can roughly be categorized in several classes that are not mutually exclusive: applying pre-processing to the input speech by means of non-ASR models \cite{investigation_hallucinations}, post-processing of text transcriptions \cite{investigation_hallucinations} or intermediate outputs of an ASR model \cite{audio_sae}, or fine-tuning specific neurons responsible for hallucinations \cite{calm_whisper}. In this paper, we consider one of the most popular ASR models called Whisper \cite{whisper} and check the hypothesis that its intermediate hidden representations can be used to predict whether the generated transcript is going to contain hallucinations. Then, we provide an efficient mechanism of fixing the hallucinations by means of steering the corresponding hidden representations or their respective activations obtained from Sparse AutoEncoder (SAE) trained on these hiddens. Through extensive experiments, we demonstrate that both raw activations and SAE latent representations encode linearly separable hallucination-related information, with classification performance improving toward deeper encoder layers. We further show that SAE-based steering consistently outperforms activation steering in hallucination reduction, achieving results competitive with fine-tuning-based methods without any modification of model parameters.

Our main contributions can be summarized as follows:
\begin{itemize}
    \item We demonstrate that hallucination-prone inputs are linearly separable from non-hallucinating ones in both raw Whisper encoder activations and SAE latent representations, with discriminative information concentrated in a sparse subset of features and increasing toward deeper encoder layers.
    \item We propose two fine-tuning-free steering strategies and show that SAE-based steering achieves consistent hallucination reduction across all evaluated non-speech datasets and both model variants, approaching the performance of fine-tuning-based methods.
    \item We provide a comprehensive empirical comparison across two Whisper model variants, multiple non-speech and speech datasets, and two languages, revealing consistent trade-offs between hallucination reduction and ASR metric preservation under steering.
\end{itemize}

\section{Background}
In this section we briefly overview Sparse AutoEncoders and steering -- two concepts we will extensively use for analysis of Whisper's hallucinations and ways of fixing them.

\subsection{Sparse AutoEncoders}
Sparse AutoEncoders (SAEs) are a class of neural network models originally developed in the context of mechanistic interpretability to decompose dense, polysemantic activation vectors into sparse, human-interpretable latent representations for NLP models \cite{sae_irish, gemma_scope}. They were further adopted to image \cite{sae_visual1, sae_visual2} and audio \cite{audio_sae, music_sae} domains. The key motivation is the superposition hypothesis: neural networks are believed to represent more features than they have dimensions by encoding multiple concepts in overlapping directions, which makes direct analysis of raw activations difficult. SAEs address this by projecting activations into a much higher-dimensional latent space while enforcing sparsity, encouraging each latent dimension to correspond to a single interpretable concept.

Formally, given an input activation $\mathbf{h} \in \mathbb{R}^{d}$, a SAE encodes it into a sparse latent representation $\mathbf{z} \in \mathbb{R}^{m}$, where $m \gg d$, and reconstructs the original activation via a linear decoder:
\[
    \mathbf{z} = \sigma(\mathbf{W}_{\mathrm{enc}}\,\mathbf{h} + \mathbf{b}_{\mathrm{enc}}), \qquad \hat{\mathbf{h}} = \mathbf{W}_{\mathrm{dec}}\,\mathbf{z} + \mathbf{b}_{\mathrm{dec}}
\]

\noindent where $\sigma(\cdot)$ denotes a sparsity-inducing activation function (e.g. ReLU or TopK). The model is trained to minimize a combination of reconstruction loss and a sparsity penalty on the latent activations:

\[
    \mathcal{L} = ||\mathbf{h} - \hat{\mathbf{h}}||_2^2 + \lambda\,\Omega(\mathbf{z})
\]

\noindent where $\lambda$ controls the trade-off between reconstruction fidelity and sparsity, and $\Omega(\mathbf{z})$ is a sparsity penalty term, typically the $L_1$ norm $||\mathbf{z}||_1$ or a related regularizer depending on the specific SAE variant employed.

\subsection{Steering}

Activation steering is a technique for intervening on a neural network's internal representations at inference time, with the goal of influencing the model output behavior in a controlled and targeted manner. The core idea is to identify a direction in the model's activation space that corresponds to a desired behavioral change, and then add a scaled version of the vector corresponding to this direction to the hidden states during the forward pass. This approach allows for direct manipulation of the model behavior without any modification to the model weights, making it a lightweight and fine-tuning-free alternative to parameter-level interventions.

Steering has been extensively studied in LLMs for sentiment control, toxicity reduction, truthfulness improvement, and suppression of undesirable behaviors \cite{turner2024steering, zou2022representation, li2023inference, rimsky2024steering}, and has been extended beyond language to diffusion models \cite{gaintseva2026casteer} and audio generation \cite{facchiano2504activation}. The steering vector is typically derived as the difference between mean activations of two contrastive input sets, a method known as Contrastive Activation Addition (CAA) \cite{rimsky2024steering}.

More recently, steering has been extended to operate in the latent space of SAEs, offering a more interpretable and disentangled alternative to raw activation steering \cite{kuznetsov2025sae}. Rather than intervening on dense, polysemantic activation vectors, SAE-based steering targets individual sparse latent dimensions that have been identified as relevant to the behavior of interest, providing finer-grained control over the model's outputs.

\section{Methodology}

\subsection{Whisper Hallucinations} \label{sec:whisper_hallucinations}

Whisper is a Transformer-based ASR model trained on a large-scale weakly supervised dataset of 680,000 hours of audio collected from the Internet. While this training paradigm enables remarkable generalization across languages, domains, and acoustic conditions, it also introduces a well-known failure mode: hallucinations. Since the training data is not manually curated, audio segments with no meaningful speech content (such as silence, background noise, or music) may be paired with arbitrary text transcriptions, leading the model to learn spurious correlations between non-speech inputs and fluent text outputs. As a result, Whisper tends to produce coherent but entirely fabricated transcriptions when presented with non-speech audio.

To mitigate this behavior, Whisper employs an internal filtering mechanism based on two scalar quantities computed during inference. The first is \texttt{no\_speech\_prob}, the probability assigned by the model to a special \texttt{<|nospeech|>} token, which serves as an explicit indicator of whether the input segment contains speech. The second is \texttt{avg\_logprob}, the average log-probability of the generated tokens, which reflects the model's overall confidence in the produced transcription. Segments are suppressed when \texttt{no\_speech\_prob} exceeds a fixed threshold and \texttt{avg\_logprob} falls below another threshold. However, this heuristic filtering is insufficient in practice: hallucinated outputs are often generated with high confidence, resulting in elevated \texttt{avg\_logprob} values, while the corresponding \texttt{no\_speech\_prob} remains unexpectedly low — both conditions allowing the hallucinated segment to pass the filter undetected. This motivates the need for more principled approaches to hallucination detection and suppression.

We define the Detection Rate (DR) as a general metric applicable to both speech and non-speech data:

\begin{equation}\label{eq:DR}
    \text{DR}(\mathcal{D}) = \frac{1}{|\mathcal{D}|} \sum_{i=1}^{|\mathcal{D}|} \mathbb{I}\{\mathrm{p}_i < \tau_p \text{ } \lor \text{ } \mathrm{l}_i > \tau_l\}
\end{equation}

\noindent where $\mathrm{p}$ and $\mathrm{l}$ denote $\mathrm{no\_speech\_prob}$ and $\mathrm{avg\_logprob}$ respectively, $\tau_p$ and $\tau_l$ are their corresponding thresholds, and $\mathbb{I}\{\cdot\}$ is the indicator function. DR quantifies the fraction of samples classified as speech-containing, and is used to derive the Hallucination Rate (HR): for non-speech data $\text{HR} = \text{DR}$, reflecting the proportion of hallucinated outputs; for speech data $\text{HR} = 1 - \text{DR}$, capturing the proportion of genuine speech samples incorrectly suppressed. We will refer to non-speech samples incorrectly identified by Whisper's internal filters as speech as hallucinating samples.

To ensure both effective hallucination reduction and reliable control over ASR performance, we collect representations from two complementary data sources. Non-speech data is used to analyze and steer the model's behavior on hallucination-prone inputs, while speech data is used to monitor the impact of our interventions on the model's transcription quality, measured via Word Error Rate (WER) for English and Character Error Rate (CER) for Chinese. This dual-data strategy allows us to optimize for hallucination suppression while preserving the model's core ASR capabilities.

\subsection{Representations}

In this work, we consider two representation spaces for hallucination analysis: raw activations extracted directly from Whisper's audio encoder, and sparse latent representations obtained from a SAE trained on those activations. The former provides direct access to the model's internal states, while the latter projects these representations into a higher-dimensional, overcomplete basis under a sparsity constraint, yielding a disentangled decomposition that may expose structure not readily apparent in the dense activation space.

\subsubsection{Whisper Activations}

We collect internal representations the residual stream, capturing the hidden states after each layer's residual addition. Formally, let $\mathbf{h}_l \in \mathbb{R}^{T \times d}$ denote the residual stream activation at layer $l$, where $T$ is the sequence length and $d$ is the model's hidden dimension. These activations are aggregated across the temporal dimension with average pooling to obtain a fixed-size representation per audio segment.

\subsubsection{SAE Representations}

We adopt the SAE training methodology of \cite{audio_sae}, which trains SAEs across all encoder layers of Whisper on a diverse corpus of audio data. The resulting sparse latent representations $\mathbf{z}$ capture acoustic, semantic, and paralinguistic information in a somewhat disentangled manner. To obtain fixed-size representations, we aggregate $\mathbf{z}$ across the temporal dimension via non-zero average pooling, computing the mean exclusively over active (non-zero) elements to preserve the sparsity structure. We use these representations as a complementary view to raw Whisper activations for both hallucination classification and steering experiments.

\subsection{Classification}

Before proposing steering strategies, we first investigate whether Whisper's internal representations encode sufficient information to distinguish hallucination-prone inputs from non-hallucinating ones. This classification task serves as a necessary diagnostic step: if the representations carry discriminative information about hallucinations, it validates the premise that targeted intervention in the activation space can meaningfully reduce hallucination prevalence. Of particular importance is whether hallucinating and non-hallucinating samples form \textit{linearly separable} clusters in the representation space, as linear separability is a prerequisite for the effectiveness of linear steering methods: a reliable linear decision boundary implies the existence of a well-defined direction in the activation space along which the model's behavior can be steered. To this end, we employ linear classifiers and evaluate their discriminative performance as a proxy for the linear structure of hallucination-related information in the representations.

Classification is performed on diverse non-speech data, where ground-truth hallucination labels are assigned per sample according to the DR rule defined in Section~\ref{sec:whisper_hallucinations}. As the primary evaluation metric, we compute the AUC Score between the probability distributions assigned by the classifier to hallucination and non-hallucination classes, providing a threshold-independent measure of separability between the two groups. For SAE latent representations, the classification step serves an additional purpose: we exploit the classifier's feature importance scores to identify the top-$k$ most discriminative SAE latent dimensions, which are subsequently used to construct the steering mask for SAE-based intervention.

\subsection{Steering}

In this work, we apply both activation space steering and SAE latent space steering to Whisper's audio encoder representations, with the goal of suppressing hallucinations on non-speech inputs while preserving transcription quality on genuine speech.

\subsubsection{Whisper Activations Steering}

Activation steering is performed by computing a steering vector $\mathbf{v} \in \mathbb{R}^{d}$ from contrastive sets of Whisper encoder activations and adding a scaled version of it to the residual stream at a selected layer $l$ during inference. The steering vector is derived as the difference between the mean activations computed over a set of hallucinating samples $\mathcal{H}$ and a set of non-hallucinating samples $\mathcal{N}$:

\[
    \mathbf{v} = \frac{1}{|\mathcal{N}|}\sum_{i \in \mathcal{N}} \mathbf{h}_l^{(i)} - \frac{1}{|\mathcal{H}|}\sum_{i \in \mathcal{H}} \mathbf{h}_l^{(i)}
\]

\noindent where $\mathbf{h}_l^{(i)} \in \mathbb{R}^{d}$ denotes the residual stream activation of sample $i$ at layer $l$. The resulting vector $\mathbf{v}$ points in the direction that shifts representations away from the hallucination-prone regime toward the non-hallucinating regime in the activation space.

At inference time, the steering intervention is applied by modifying the residual stream at layer $l$ for each input token position $t$:

\[
    \tilde{\mathbf{h}}_l^{(t)} = \mathbf{h}_l^{(t)} + \alpha \cdot \mathbf{v}
\]

\noindent where $\alpha \in \mathbb{R}$ is a scalar coefficient controlling the magnitude of the intervention, and $\mathbf{v}$ is the steering vector. A positive $\alpha$ steers the representations toward the non-hallucinating direction, while the magnitude of $\alpha$ determines the strength of the intervention. The modified activation $\tilde{\mathbf{h}}_l^{(t)}$ is then passed to the subsequent Transformer block, propagating the steering effect through the remainder of the encoder.

\subsubsection{SAE Representations Steering} \label{sec:sae_steering}

SAE-based steering operates directly in the sparse latent space rather than in the dense activation space. Given a sparse latent representation $\mathbf{z} \in \mathbb{R}^{m}$ obtained from the SAE encoder, we intervene on the latent dimensions identified as most discriminative by the classifier, and then reconstruct the steered activation via the SAE decoder for injection back into the residual stream.

Let $\boldsymbol{\beta} \in \mathbb{R}^{m}$ denote the vector of feature importance scores obtained from the hallucination classifier, where positive values of $\beta_j$ indicate that the $j$-th latent dimension contributes positively to the hallucination label. Since hallucinating samples are labeled as 1 and non-hallucinating samples as 0, a positive importance score corresponds to a feature that increases hallucination probability. We select the top-$k$ most representative features based on the absolute values of $\boldsymbol{\beta}$, retaining only the $k$ dimensions with the largest $|\beta_j|$. Let $\boldsymbol{\beta}_{\text{topk}}$ denote the sparse vector containing these $k$ values and zeros elsewhere. To steer away from hallucinations, we define the steering direction as:

\[
    \mathbf{s} = -\text{sign}(\boldsymbol{\beta}_{\text{topk}}) \in \{-1, 0, 1\}^{m}
\]

\noindent where $\mathbf{s}$ is a sparse direction vector retaining only $k$ non-zero entries, with the sign flipped to intervene against the hallucination-promoting direction.

We further define $\bar{\mathbf{Z}} \in \mathbb{R}^{m}$ as the vector of average SAE latent activations computed over a reference dataset, which provides a data-driven estimate of the typical activation magnitude per latent dimension and is used to calibrate the scale of the intervention.

We propose and compare two steering strategies. The \textit{additive} method shifts the latent representation along the steering direction by a scaled offset:

\[
    \mathbf{z}' = \mathbf{z} + \alpha \cdot \mathbf{s} \odot \bar{\mathbf{Z}}
\]

\noindent where $\alpha \in \mathbb{R}$ is a scalar coefficient controlling the intervention strength, and $\odot$ denotes element-wise multiplication. This formulation ensures that the magnitude of the additive shift is proportional to the typical activation scale of each latent dimension, preventing disproportionate perturbations in dimensions with naturally small activations.

The \textit{multiplicative} method instead scales the latent representation along the steering direction:

\[
    \mathbf{z}' = \mathbf{z} \odot \alpha^{\mathbf{s}}
\]

\noindent where $\alpha^{\mathbf{s}}$ denotes element-wise exponentiation of the scalar $\alpha$ by the entries of $\mathbf{s}$. Since $\mathbf{s} \in \{-1, 0, 1\}^{m}$, this operation yields $1/\alpha$ for dimensions where $s_j = -1$ (suppressing hallucination-promoting features), $\alpha$ for dimensions where $s_j = 1$ (amplifying non-hallucination-promoting features), and $1$ for dimensions where $s_j = 0$ (leaving unselected dimensions unchanged). The steered latent $\mathbf{z}'$ is then passed through the SAE decoder to reconstruct the modified activation $\hat{\mathbf{h}}' = \mathbf{W}_{\mathrm{dec}}\,\mathbf{z}' + \mathbf{b}_{\mathrm{dec}}$, which is subsequently injected into the residual stream in place of the original encoder activation.

\section{Experimental Setup}


\subsection{Models}

We conduct all experiments on two Whisper model variants: Whisper small and Whisper large-v3. This choice is motivated by the desire to evaluate our proposed methods across models of substantially different scales, allowing us to assess whether the observed effects generalize beyond a single model size.

For inference, we use greedy decoding with default Whisper decoding options, setting temperature to zero and disabling beam search.

For SAE inference, we use the corresponding AudioSAE Hugging Face checkpoints:
Whisper small SAE\footnote{\url{https://huggingface.co/Egorgij21/Audio-SAE-Whisper-small}}
and Whisper large-v3 SAE\footnote{\url{https://huggingface.co/Egorgij21/Audio-SAE-Whisper-large-v3}}.
We use the Batch-Top-$k$ SAE architecture implementation released in the GitHub repository\footnote{\url{https://github.com/audiosae/audio-sae}}.

The SAE checkpoints use an expansion coefficient of 8 and $k = 50$.
The expansion coefficient defines the ratio between the SAE latent dimension $m$ and the encoder hidden dimension $d$, yielding latent spaces of dimension $m = 8 \times 768 = 6144$ and $m = 8 \times 1280 = 10240$ for Whisper small and large-v3 respectively. The Top-$k$ parameter $k = 50$ enforces that exactly 50 latent dimensions are active per token, providing a fixed and consistent sparsity level across all inputs and model variants.


\subsection{Dataset}

Our experimental setup relies on two different types of data: non-speech data, used for hallucination analysis, classifier training and steering vector calculation, and speech data, used for monitoring ASR metric quality. The dataset splits are strictly defined to prevent any leakage between training and evaluation: all classifiers are trained, steering vectors are computed, and steering hyperparameters are tuned solely on the non-speech train split, with ASR metrics monitored on speech data. The non-speech test split is reserved exclusively for final evaluation to assess the generalization of the proposed methods to unseen non-speech domains. A complete overview of all datasets used in our experiments is provided in Table~\ref{tab:datasets}.

\subsubsection{Non-Speech Data}
The non-speech train split comprises three datasets covering a wide variety of acoustic conditions: MUSAN \cite{musan} noise, WHAM! \cite{wham} training subset, and the development subset of FSD50k \cite{fsd50k}. The non-speech test split consists of UrbanSound8K \cite{urbansound8k}, the validation and test subsets of WHAM!, and the evaluation subset of FSD50k. To ensure that no speech-like content contaminates the non-speech experiments, we additionally define FSD50k-\textit{filtered} for both splits as the subset of FSD50k obtained by removing all samples annotated with labels containing \textit{speech}, \textit{music}, or \textit{human} categories, resulting in 3,463 retained samples in both train and test splits. We further define \textit{FULL} train and test datasets as the union of all samples from the respective non-speech splits, where the unfiltered version of FSD50k is used. \textit{Full} datasets contain 61896 and 26963 samples for train and test respectively. 

\subsubsection{Speech Data}
All speech data is used to evaluate the impact of our steering interventions on ASR performance. For English, we use LibriSpeech \cite{librispeech} test-clean and test-other and the English subset of FLEURS \cite{fleurs}. For Chinese, we use the Chinese subset of FLEURS and AISHELL-1 \cite{aishell1}. We evaluate transcription quality via WER and CER for English and Chinese respectively.

\begin{table}[t]
    \centering
    \caption{Overview of datasets used in experiments.}
    \label{tab:datasets}
    \resizebox{\columnwidth}{!}{%
    \begin{tabular}{lllll}
        \toprule
        \textbf{Dataset} & \textbf{Subset} & \textbf{Type} & \textbf{Split} & \textbf{N Samples} \\
        \midrule
        MUSAN Noise & noise & & & 930 \\
        WHAM! & tr & & & 20000 \\
        FSD50k & dev & Non-speech & Train & 40966 \\
        FSD50k-filtered & dev & & & 38607 \\
        FULL & — & & & 61896 \\
        \midrule
        UrbanSound8K & — & & & 8732 \\
        WHAM! & cv / tt & & & 8000 \\
        FSD50k & eval & Non-speech & Test & 10231 \\
        FSD50k-filtered & eval & & & 9127 \\
        FULL & — & & & 26963 \\
        \midrule
        LibriSpeech & test-clean & Speech (EN) & & 2620 \\
        LibriSpeech  & test-other & Speech (EN) & & 2939 \\
        FLEURS & en & Speech (EN) & Test & 647 \\
        FLEURS  & zh & Speech (ZH) & & 945 \\
        AISHELL-1 & — & Speech (ZH) & & 141600 \\
        \bottomrule
    \end{tabular}%
    }
\end{table}

\subsubsection{Distributions}

To gain insight into the hallucination behavior of both Whisper variants across the collected datasets, we analyze the joint distribution of the two internal filtering parameters: $\mathrm{avg\_logprob}$ and $\mathrm{no\_speech\_prob}$. Following the default thresholds used in the original Whisper inference code, we set $\tau_l = -1.0$ for $\mathrm{avg\_logprob}$ and $\tau_p = 0.6$ for $\mathrm{no\_speech\_prob}$, defining the hallucination region as samples satisfying Equation~\ref{eq:DR} for non-speech data, and the opposite for speech.


A quantitative summary of hallucination rates per dataset and model is provided in Table~\ref{tab:hallu_rates}. The results confirm that hallucination prevalence varies substantially across datasets and model sizes.

\begin{table}[t]
    \centering
    \caption{Hallucination Rate (HR, \%) per dataset and model variant, computed on train and test splits of non-speech data.}
    \label{tab:hallu_rates}
    \resizebox{\columnwidth}{!}{%
    \begin{tabular}{lcccc}
        \toprule
        \multirow{2}{*}{\textbf{Dataset}} & \multicolumn{2}{c}{\textbf{Whisper small}} & \multicolumn{2}{c}{\textbf{Whisper large-v3}} \\
        \cmidrule(lr){2-3} \cmidrule(lr){4-5}
        & \textbf{Train HR (\%)} & \textbf{Test HR (\%)} & \textbf{Train HR (\%)} & \textbf{Test HR (\%)} \\
        \midrule
        MUSAN Noise      & 79.2 & — & 63.1 & — \\
        WHAM!            & 68.9 & 66.9 & 93.1 & 92.0 \\
        FSD50k           & 80.6 & 81.9 & 69.0 & 75.0 \\
        FSD50k-filtered  & 80.2 & 80.8 & 68.5 & 73.6 \\
        UrbanSound8k     & — & 67.0 & — & 95.2 \\
        FULL             & 76.8 & 72.6 & 76.7 & 86.8 \\
        \bottomrule
    \end{tabular}%
    }
\end{table}

\subsection{Classification} \label{sec:classification_setup}

For classification task, hallucination labels are assigned to each sample according to the DR rule defined in Section~\ref{sec:whisper_hallucinations}. Activations are extracted independently for each encoder layer, yielding a separate classification problem per layer.

We train a logistic regression classifier from \texttt{scikit-learn} library with the \textit{saga} solver for each layer independently, preceded by a \texttt{MaxAbsScaler} normalization step. To obtain reliable performance estimates, we employ stratified 5-fold cross-validation on each train dataset, ensuring balanced class proportions across folds. At each fold, the scaler is fit exclusively on the training portion and applied to both the validation fold and all test datasets. Fold-level AUC scores are averaged to report mean and standard deviation on both validation and test sets.

SAE latent representations follow the same cross-validation protocol. The logistic regression classifier is applied to the full SAE latent vector, and the resulting feature importance coefficients $\boldsymbol{\beta}$ are collected across all folds and averaged to produce estimates of per-feature discriminative power. These aggregated importance scores are subsequently used to identify the top-$k$ most relevant SAE latent dimensions for steering, as described in Section~\ref{sec:sae_steering}.

\subsection{Steering}

Both steering strategies share a hyperparameter tuning protocol. All hyperparameters are optimized via grid search exclusively on non-speech train data, with the HR metric, defined in Section~\ref{sec:whisper_hallucinations}, as the primary optimization objective. While WER and CER on speech data are monitored concurrently to ensure that transcription quality is not severely degraded. Non-speech test data is strictly excluded from hyperparameter tuning and reserved for final evaluation only. We compute WER aggregated across all English speech datasets and CER aggregated across all Chinese speech datasets.

For activation steering, the hyperparameter is the steering coefficient $\alpha$, controlling the magnitude of the intervention. For SAE-based steering, two hyperparameters are optimized jointly: the steering coefficient $\alpha$ and the number of intervened latent dimensions $k$, which controls the sparsity of the steering mask $\mathbf{s}$. Both additive and multiplicative SAE steering variants are evaluated under this same tuning procedure.

For both steering strategies, intervention is applied at the final encoder layer of each model, which is identified as the most discriminative layer based on classification results. This single-layer configuration avoids potentially conflicting perturbations across multiple layers and simplifies the overall intervention scheme.

\section{Experimental Results}

\subsection{Classification}

\begin{figure*}[t]
    \centering
    \includegraphics[width=0.49\textwidth]{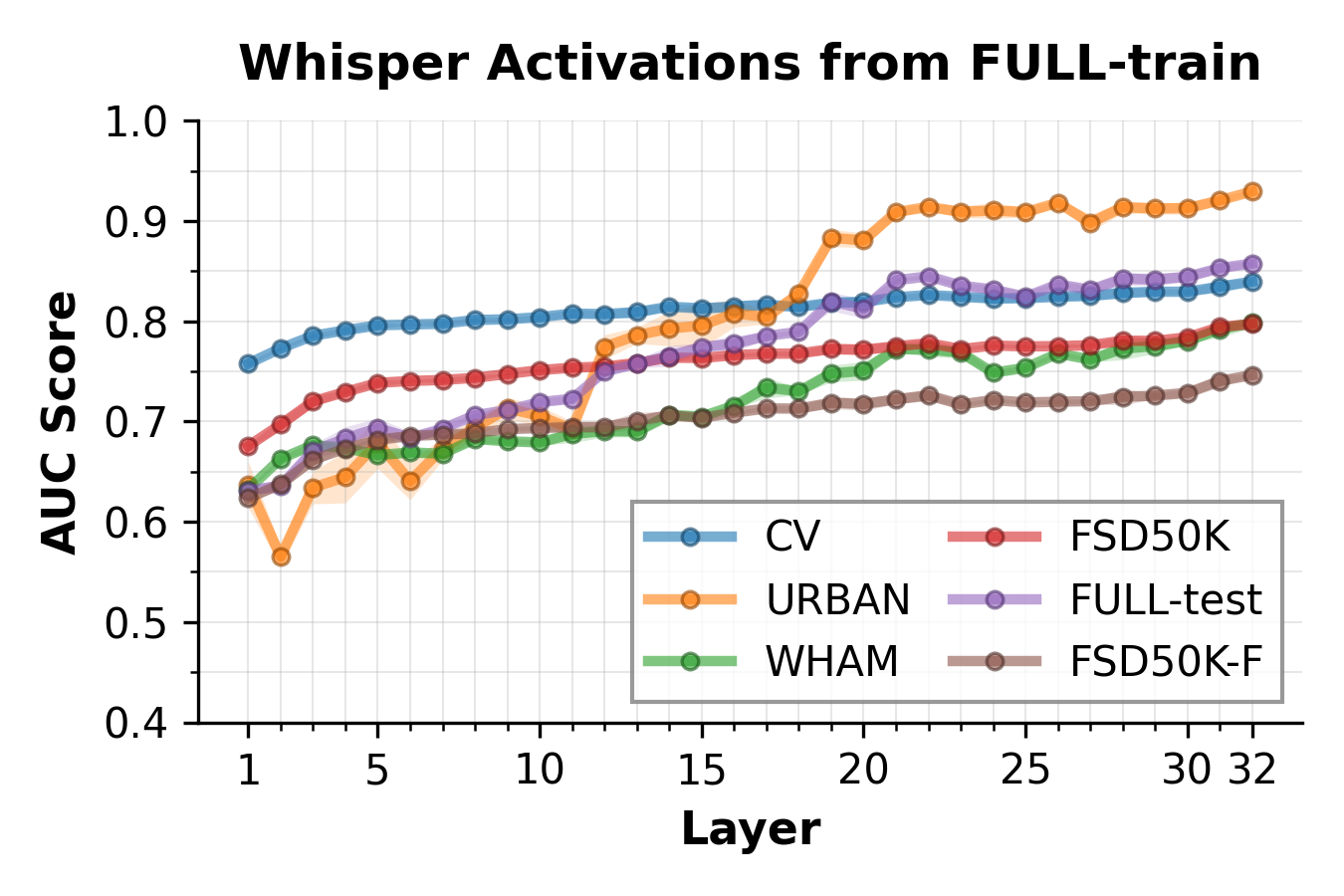}
    \hfill
    \includegraphics[width=0.49\textwidth]{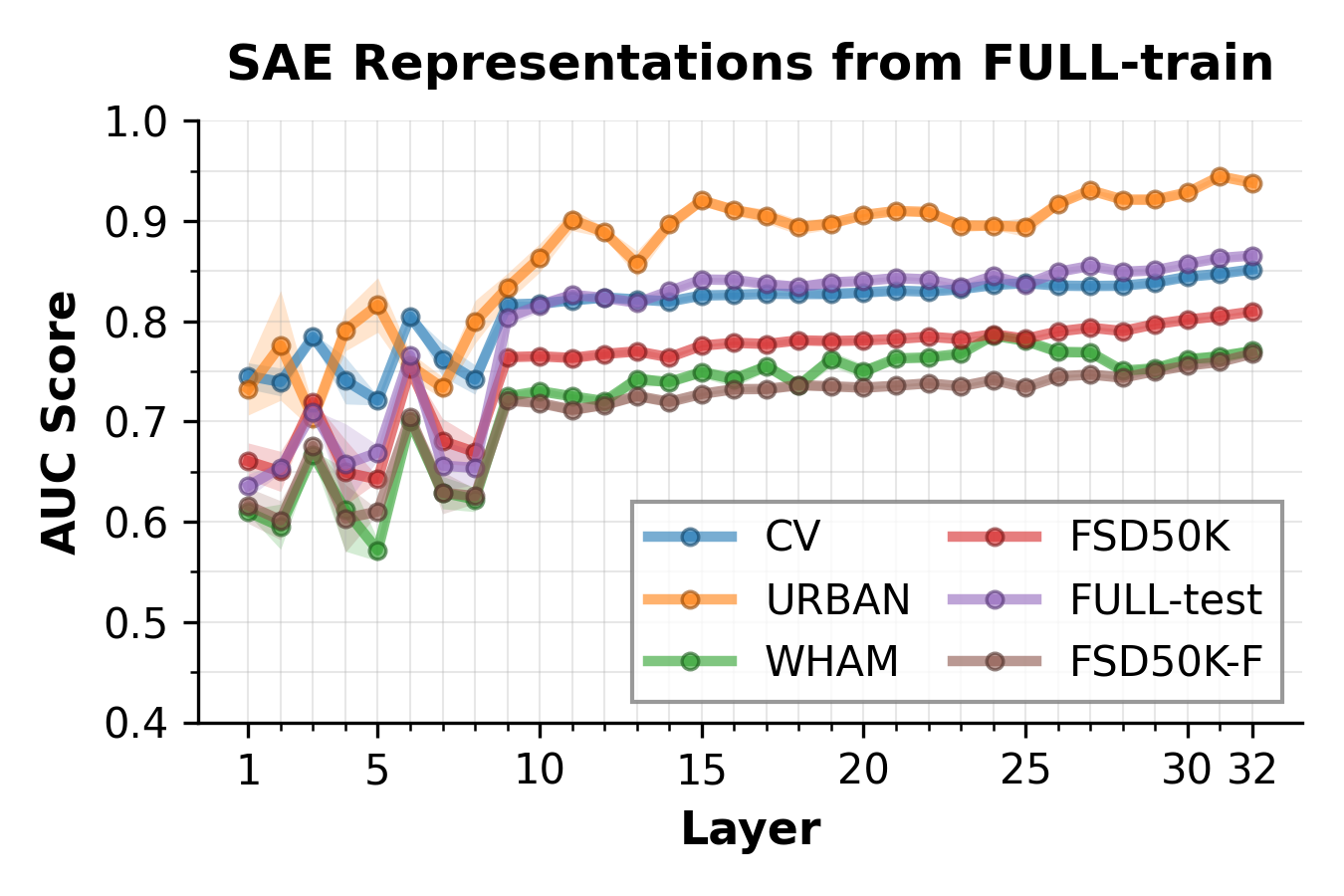}
    \caption{Layer-wise AUC scores for classifiers trained on raw Whisper activations (left) and SAE latent representations (right) for Whisper large-v3. Each curve corresponds to a test dataset, except for the CV curve, which displays the cross-validation score on the training set. All results correspond to training on the FULL-train dataset.}
    \label{fig:clf_whisper}
\end{figure*}

\subsubsection{Whisper Activations}

Figure~\ref{fig:clf_whisper} (left column) presents the layer-wise AUC scores obtained by the logistic regression classifier trained on raw Whisper encoder activations for Whisper large-v3. A trend is observed across all train and test datasets: classification performance improves with layer depth, with the highest AUC scores concentrated in the final encoder layers. This suggests that hallucination-related information becomes more linearly separable as representations propagate through the encoder, with deeper layers encoding more abstract and discriminative features. This observation motivates the choice of the final encoder layer as the intervention point for activation steering.

\subsubsection{SAE Representations}

Figure~\ref{fig:clf_whisper} (right column) presents the analogous layer-wise AUC scores obtained on SAE latent representations. The same trend of improving classification performance with layer depth is observed. Notably, SAE representations achieve competitive or superior AUC scores on several datasets. This suggests that sparse representations preserve, and in some cases enhance, the discriminative structure present in the original activations.

\begin{figure}[t]
    \centering
    \includegraphics[width=\columnwidth]{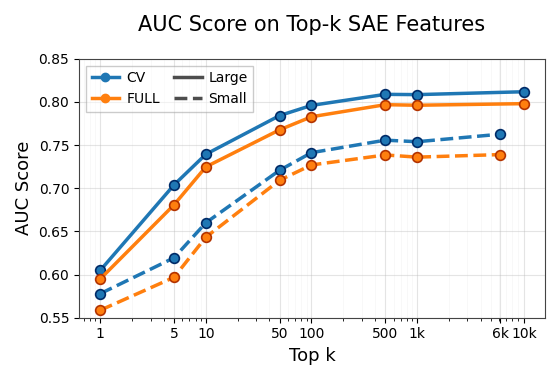}
    \caption{AUC score against the number of top-$k$ SAE features for Whisper small and large-v3, with cross-validation (CV) and FULL-test curves shown for both models, averaged by layers.}
    \label{fig:clf_topk}
\end{figure}

Figure~\ref{fig:clf_topk} shows the classification AUC as a function of the number of top-$k$ SAE features used. For both Whisper small and large-v3, performance stabilizes at 50--100 features, with little gain observed when adding more dimensions. This indicates that hallucination-related information is concentrated in a small number of SAE latent dimensions, which supports the use of a steering vector with a with a small number of non-zero indices in subsequent experiments.

\subsubsection{Comparison}

Table~\ref{tab:clf_comparison} summarizes the classification results averaged across all encoder layers. Both representation spaces consistently achieve above-chance AUC scores across all datasets and model variants, confirming that hallucination-related information is encoded in both raw Whisper activations and SAE latent representations. SAE representations show a slight advantage for Whisper large-v3, while results for Whisper small are comparable across both spaces. Across both models and representation types, the FULL-train dataset consistently yields the best or near-best AUC scores on the FULL-test set, demonstrating the benefit of diverse training data for generalization. Based on these results, we select the FULL-train dataset for computing steering vectors and SAE feature importance scores in all subsequent steering experiments.

\begin{table*}[t]
    \centering
    \caption{Layer-averaged AUC scores for logistic regression classifiers trained on raw Whisper activations and SAE latent representations, for Whisper small and large-v3 across all datasets. CV denotes cross-validation score on the train dataset.}
    \label{tab:clf_comparison}
    \resizebox{\textwidth}{!}{%
    \begin{tabular}{llccccccc}
        \toprule
        \multirow{2}{*}{\textbf{Repr.}} & \multirow{2}{*}{\textbf{Model}} & \multirow{2}{*}{\textbf{Train Dataset}} & \multirow{2}{*}{\textbf{CV AUC}} & \multicolumn{5}{c}{\textbf{Test AUC}} \\
        \cmidrule(lr){5-9}
        & & & & \textbf{UrbanSound8K} & \textbf{WHAM! cv/tt} & \textbf{FSD50k eval} & \textbf{FSD50k-filtered eval} & \textbf{FULL test} \\
        \midrule
        \multirow{6}{*}{Whisper} 
        & \multirow{3}{*}{small}   & Musan noise & 0.75 & 0.57 & 0.46 & 0.66 & 0.69 & 0.57 \\
        &                          & WHAM! tr    & 0.77 & 0.62 & \textbf{0.77} & 0.64 & 0.65 & 0.69 \\
        &                          & FSD50k dev  & 0.79 & 0.66 & 0.62 & \textbf{0.78} & \textbf{0.77} & 0.67 \\
        &                          & FULL train & 0.77 & \textbf{0.71} & 0.75 & 0.77 & \textbf{0.77} & \textbf{0.76} \\
        \cmidrule(lr){2-9}
        & \multirow{3}{*}{large-v3} & Musan noise & 0.68 & 0.57 & 0.56 & 0.57 & 0.55 & 0.62 \\
        &                           & WHAM! tr    & 0.79 & 0.74 & \textbf{0.78} & 0.59 & 0.56 & 0.63 \\
        &                           & FSD50k dev  & 0.77 & \textbf{0.80} & 0.72 & \textbf{0.76} & \textbf{0.70} & \textbf{0.77} \\
        &                          & FULL train & 0.73 & \textbf{0.80} & 0.72 & \textbf{0.76} & \textbf{0.70} & \textbf{0.77} \\
        \midrule
        \multirow{6}{*}{SAE}
        & \multirow{3}{*}{small}   & Musan noise & 0.72 & 0.58 & 0.49 & 0.69 & 0.71 & 0.58 \\
        &                          & WHAM! tr    & 0.73 & 0.61 & \textbf{0.73} & 0.62 & 0.62 & 0.66 \\
        &                          & FSD50k dev  & 0.79 & 0.66 & 0.58 & \textbf{0.78} & \textbf{0.78} & 0.66 \\
        &                          & FULL train & 0.76 & \textbf{0.68} & \textbf{0.73} & 0.77 & 0.76 & \textbf{0.74} \\
        \cmidrule(lr){2-9}
        & \multirow{3}{*}{large-v3} & Musan noise & 0.68 & 0.64 & 0.55 & 0.61 & 0.60 & 0.65 \\
        &                           & WHAM! tr    & 0.76 & 0.75 & \textbf{0.73} & 0.59 & 0.55 & 0.68 \\
        &                           & FSD50k dev  & 0.77 & 0.85 & 0.62 & \textbf{0.76} & \textbf{0.71} & 0.78 \\
        &                          & FULL train & 0.81 & \textbf{0.87} & 0.72 & \textbf{0.76} & \textbf{0.71} & \textbf{0.80} \\
        \bottomrule
    \end{tabular}%
    }
\end{table*}

\subsection{Steering}

Based on the classification results, all steering hyperparameters are tuned using activations from the final layer, which consistently demonstrated the highest discriminative performance. Although the optimal steering coefficient $\alpha$ is not guaranteed to transfer between layers, we adopt this single-layer tuning strategy for simplicity and apply the resulting hyperparameters uniformly across all layers in subsequent steering experiments. This allows us to evaluate the contribution of each encoder layer to hallucination reduction under a shared hyperparameter configuration. 

\subsubsection{Whisper Activations}

\begin{figure}[t]
    \centering
    \includegraphics[width=\columnwidth]{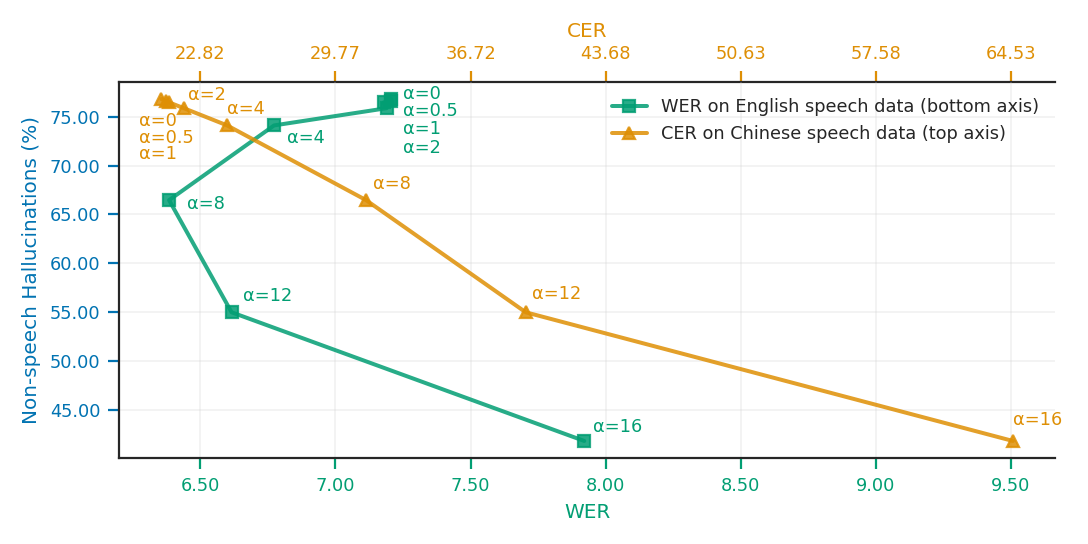}
    \\[\smallskipamount]
    \includegraphics[width=\columnwidth]{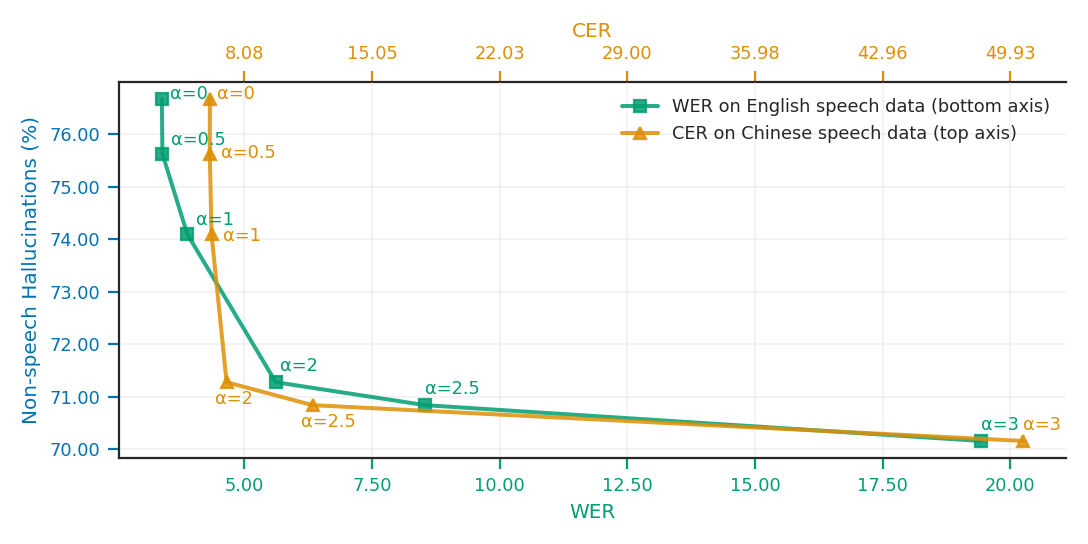}
    \caption{Trade-off between HR (left axis) and ASR quality (WER bottom, CER top) for activation steering across $\alpha$ values, for Whisper small (top) and Whisper large-v3 (bottom).}
    \label{fig:steering_whisper_pareto}
\end{figure}

Figure~\ref{fig:steering_whisper_pareto} presents the hyperparameter tuning results for activation steering, showing the trade-off between hallucination reduction (HR on FULL-train, left axis) and ASR metric degradation (WER on bottom, CER on top) as a function of $\alpha$. The Pareto frontier reveals that for Whisper small, $\alpha = 8$ provides the best balance between hallucination reduction and transcription quality preservation. For Whisper large-v3, a considerably smaller value of $\alpha = 2$ is sufficient.

\subsubsection{SAE Representations}

\begin{figure}[t]
    \centering
    \includegraphics[width=\columnwidth]{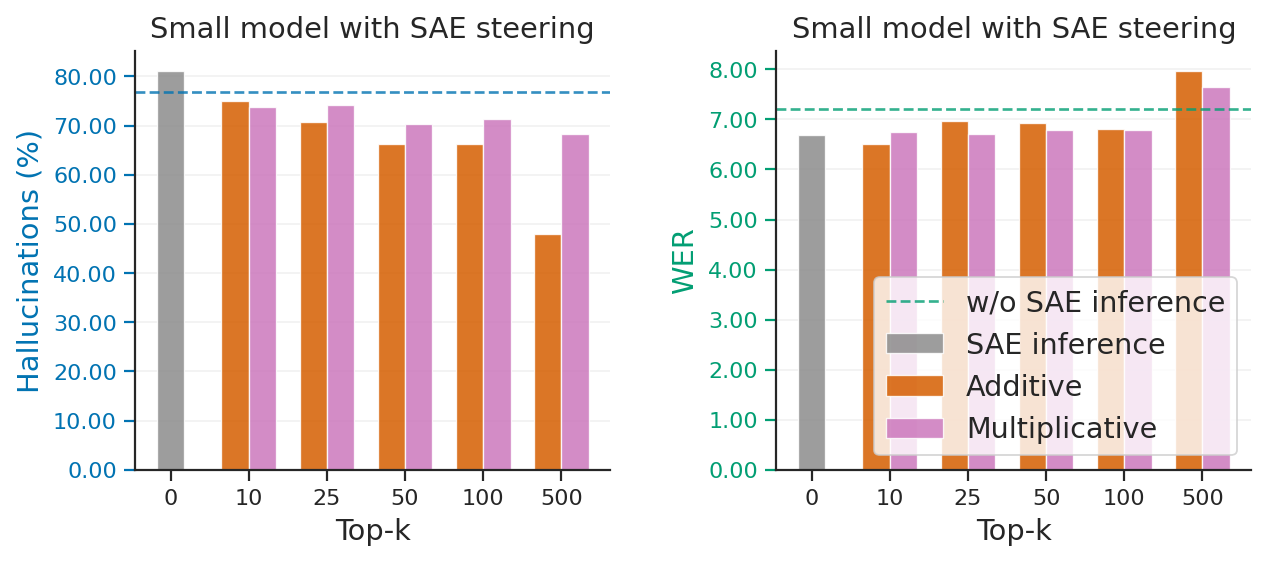}
    \caption{Comparison of additive and multiplicative SAE steering across top-$k$ values for Whisper small and large-v3, in terms of HR and WER ($\alpha=0.5$ and $\alpha=1.5$ respectively). $\alpha = 0$ denotes inference without steering.}
    \label{fig:steering_sae_barplot}
\end{figure}

\begin{figure}[t]
    \centering
    \includegraphics[width=\columnwidth]{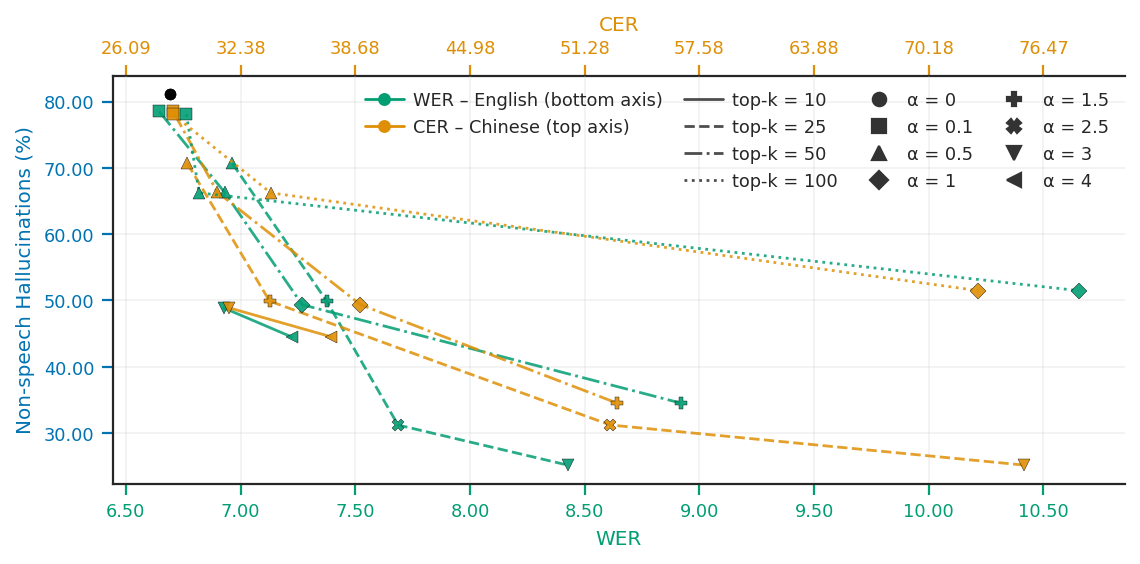}
    \\[\smallskipamount]
    \includegraphics[width=\columnwidth]{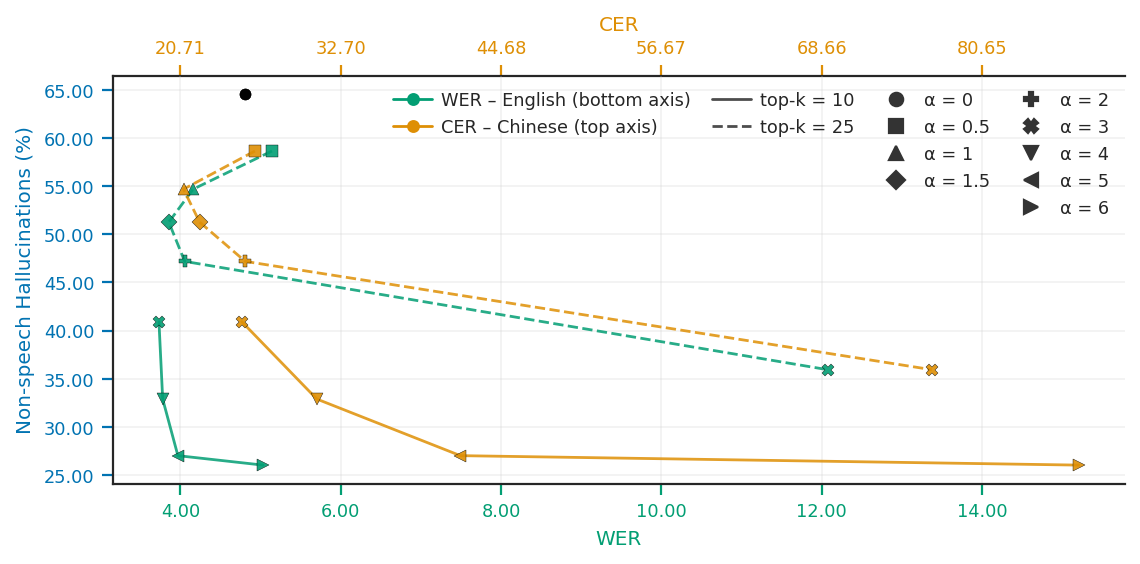}
    \caption{Trade-off between HR, CER and WER for SAE-based additive steering for small (top) and large-v3 (bottom).}
    
    \label{fig:steering_sae_pareto}
\end{figure}

Figure~\ref{fig:steering_sae_barplot} presents the bar plot comparing additive and multiplicative SAE steering configurations across different values of top-$k$ for both model variants. For the top-$k$ comparison, we fix $\alpha = 0.5$ for the additive method, corresponding to adding or subtracting half of the average activation magnitude computed on FULL-train, and $\alpha = 1.5$ for the multiplicative method, which either scales the feature by $1.5$ or suppresses it by $1.5^{-1}$ depending on the steering direction. For Whisper small, the additive method with smaller values of $k$ demonstrates better hallucination reduction in terms of HR, while the multiplicative configuration shows an advantage at larger $k$ values in terms of WER preservation. This behaviour arises from a property of the multiplicative method: if a given SAE feature is zero for a particular token, the multiplicative intervention leaves it unchanged regardless of its importance weight, limiting the effective reach of the steering signal. The additive method does not suffer from this limitation, as it introduces a non-zero offset even for inactive features. Based on these observations, we adopt the additive configuration throughout and restrict top-$k$ to small values.

The advantage of additive steering at small $k$ is consistent with the sparsity of SAE representations. Multiplicative steering can only rescale features that are already active, whereas additive steering can also modify dimensions that are inactive for a given input but identified by the classifier as discriminative. This makes additive steering more effective for non-speech inputs, where hallucinations may depend on a small set of misactivated or missing latent features.

Figure~\ref{fig:steering_sae_pareto} presents the Pareto trade-off curves for SAE-based additive steering. For Whisper small, $\alpha = 3$ with top-$k = 25$ provides the optimal balance between hallucination reduction and transcription quality preservation. For Whisper large-v3, $\alpha = 5$ with top-$k = 10$ is selected.

The strong performance obtained with small top-$k$ values suggests that the hallucination-related signal in SAE space is concentrated in a limited subset of sparse features rather than being uniformly distributed across the representation. This is consistent with the classifier results, where using only the highest-weighted SAE dimensions preserves most of the hallucination-detection performance. Steering only this small subset is therefore likely to reduce hallucinations while limiting unnecessary perturbations to the rest of the encoder representation, which may explain why SAE steering achieves a favorable trade-off between hallucination reduction and ASR quality.

A notable observation is that for Whisper small, both activation and SAE steering consistently improve WER on English data, suggesting that the steering intervention may regularize speech representations by suppressing hallucination-prone directions and shifting activations toward a more speech-discriminative region of the representation space. No such effect is observed for Whisper large-v3, where ASR metrics remain stable or exhibit marginal degradation. At the same time, CER on Chinese data increases across both steering strategies, indicating that this effect is not language-neutral. SAE inference without any steering already significantly degrades CER, likely because the SAE was not trained on Chinese speech data, making Chinese speech out-of-domain for the SAE and causing its latent reconstruction or intervention to distort representations needed for Mandarin recognition.

\subsubsection{Comparison}

Table~\ref{tab:steering_comparison} presents the best steering results for both activation and SAE-based approaches, reporting HR on all non-speech test datasets and WER/CER on speech datasets. The results demonstrate that SAE-based steering consistently achieves greater hallucination reduction than activation steering across both model variants and all non-speech test datasets.



\begin{table*}[t]
    \centering
    \caption{Comparison of activation and SAE-based steering results for Whisper small and Whisper large-v3. HR (\%) is reported for all non-speech test datasets. WER and CER are reported for English and Chinese speech datasets respectively. Baseline denotes the unsteered model. "+" in Layer column when steering is sequentially applied to those layers.}
    \label{tab:steering_comparison}
    \resizebox{\textwidth}{!}{%
    \begin{tabular}{llccccccccccccc}
        \toprule
        \multirow{2}{*}{\textbf{Repr.}} & \multirow{2}{*}{\textbf{Model}} & \multirow{2}{*}{\textbf{Layer}} & \multirow{2}{*}{\textbf{$\alpha$}} & \multirow{2}{*}{\textbf{top-$k$}} & \multicolumn{4}{c}{\textbf{HR (\%) $\downarrow$}} & \multicolumn{3}{c}{\textbf{WER $\downarrow$} (En)} & \multicolumn{2}{c}{\textbf{CER $\downarrow$} (Zh)} \\
        \cmidrule(lr){6-9} \cmidrule(lr){10-12} \cmidrule(lr){13-14}
        & & & & & \textbf{Urban} & \textbf{WHAM} & \textbf{FSD50k} & \textbf{FULL} & \textbf{LS clean} & \textbf{LS other} & \textbf{Fleurs} & \textbf{Fleurs} & \textbf{AISHELL-1} \\
        \midrule
        \multirow{11}{*}{Whisper}
        & & — & 0 & — & 67.09 & 66.93 & 81.95 & 72.63 & \textbf{3.50} & 10.28 & 9.60 & 21.96 & 26.98 \\
        & & 6 & 8 & — & 66.90 & 58.86 & 81.25 & 69.96 & 3.68 & 11.36 & \textbf{8.18} & \textbf{20.97} & \textbf{24.29} \\
        & & 9 & 8 & — & 63.53 & 47.46 & 74.12 & 62.78 & 3.83 & 9.33 & 8.58 & 24.81 & 28.84  \\
        & small & 12 & 8 & — & 66.08 & 54.93 & 73.34 & 65.55 & 3.53 & \textbf{8.75} & 8.22 & 32.25 & 32.95 \\
        & & 6+9 & 8 & — & 60.35 & 36.73 & 69.98 & 57.00 & 4.73 & 10.15 & 9.70 & 25.32 & 28.00 \\
        & & 9+12 & 8 & — & 50.85 & 29.81 & 56.79 & 46.86 & 4.07 & 8.95 & 9.27 & 38.53 & 37.15 \\
        & & 6+9+12 & 8 & — & \textbf{46.21} & \textbf{22.29} & \textbf{50.13} & \textbf{40.60} & 4.98 & 10.76 & 10.34 & 62.38 & 39.67 \\
        & & avg & 8 & — & 66.06 & 58.67 & 78.19 & 68.47 & 3.64 & 10.09 & 8.36 & 24.24 & 28.25 \\
        \cmidrule(lr){2-14}
        & & — & 0 & — & 95.98 & 92.04 & 75.08 & 86.88 & \textbf{2.11} & \textbf{4.02} & \textbf{5.75} & 9.29 & \textbf{10.54} \\
        & large-v3 & 32 & 2 & — & \textbf{90.21} & \textbf{89.46} & \textbf{70.12} & \textbf{82.37} & 4.11 & 6.73 & 7.09 & \textbf{5.62} & 14.21 \\
        & & avg & 2 & — & 95.74 & 91.21 & 74.89 & 86.49 & 2.23 & 4.17 & 5.83 & 9.46 & 10.74 \\

        \midrule
        \multirow{10}{*}{SAE}
        & & avg & 0 & 0 & 67.30 & 66.79 & 81.70 & 72.61 & \textbf{3.67} & \textbf{8.94} & \textbf{8.81} & \textbf{30.43} & \textbf{36.04} \\
        & & 1 & 3 & 25 & \textbf{0.00} & \textbf{0.03} & \textbf{2.33} & \textbf{0.89} & 123.95 & 114.66 & 132.64 & 750.76 & 127.91 \\
        & small    & 11 & 3 & 25 & 51.91 & 22.10 & 38.75 & 38.07 & 4.12 & 9.06 & 9.48 & 33.57 & 35.70 \\
        & & 12 & 3 & 25 & \underline{8.68} & \underline{4.68} & \underline{26.12} & \underline{14.11} & 4.45 & 11.34 & 12.38 & 79.01 & 92.98 \\
        & & avg & 3 & 25 & 37.94 & 31.84 & 42.17 & 37.73 & 86.45 & 98.15 & 77.07 & 214.60 & 290.30 \\
        \cmidrule(lr){2-14}
        & & avg & 0 & 0 & 78.98 & 84.92 & 73.67 & 78.73 & 6.55 & 10.06 & 11.55 & 19.09 & 21.05 \\
        & large-v3 & 26 & 5 & 10 & 50.42 & 54.74 & 60.33 & 55.46 & \textbf{2.27} & \textbf{4.21} & \textbf{6.05} & \textbf{21.47} & \textbf{26.61} \\
        & & 32 & 5 & 10 & 30.68 & \textbf{25.13} & \textbf{30.68} & 29.03 & 2.38 & 4.87 & 6.55 & 42.87 & 39.27 \\
        & & 26+32 & 3.5 & 10 & \textbf{19.88} & 27.05 & 33.92 & \textbf{27.33} & 3.70 & 7.58 & 8.73 & 59.92 & 66.24 \\
        & & avg & 5 & 10 & 69.31 & 79.02 & 67.97 & 71.68 & 12.60 & 15.41 & 16.21 & 32.42 & 39.70 \\
        \bottomrule
    \end{tabular}%
    }
\end{table*}

\subsection{Reduction of Hallucinations}

To contextualise our results, we compare our approach against Calm-Whisper \cite{calm_whisper}, which represents the closest alternative targeting the same problem. Calm-Whisper identifies a small subset of self-attention heads in the Whisper large-v3 decoder that are disproportionately responsible for hallucinations by performing head-wise masking on UrbanSound8K. Prior to fine-tuning, Calm-Whisper also reports results obtained by zeroing out the identified hallucinatory heads without any parameter updates, providing a fine-tuning-free baseline more directly comparable to our approach. Unlike our method, Calm-Whisper evaluated exclusively on Whisper large-v3.

This comparison also reflects a difference in intervention location. Calm-Whisper modifies decoder behavior, whereas our method steers encoder representations before decoding. The competitive performance of encoder-side SAE steering suggests that hallucinations are not only a decoder-level generation issue, but are already reflected in the encoder representations produced for non-speech inputs.

Table~\ref{tab:hallu_comparison} compares all methods. For comparison with Calm-Whisper, we report HR on UrbanSound8K and WER on LibriSpeech test clean and other.

\begin{table}[t]
    \centering
    \caption{Comparison of methods for Whisper large-v3. HR (\%) is reported on UrbanSound8K and on the FULL test set. WER is reported on LibriSpeech test-clean and test-other.}
    \label{tab:hallu_comparison}
    \resizebox{\columnwidth}{!}{%
    \begin{tabular}{llccc}
        \toprule
        \textbf{Method} & \textbf{HR (\%) $\downarrow$} & \textbf{clean WER $\downarrow$} & \textbf{other WER $\downarrow$} \\
        \midrule
        Baseline                                                & 95.98 & \textbf{2.11} & \textbf{4.02} \\
        Calm-Whisper (masking) & 24.10 & 3.57 & 5.98 \\
        Calm-Whisper (fine-tuned) & \textbf{15.51} & 2.19 & 4.13 \\
        \midrule
        Activation steering                              & 90.21 & 4.11 & 6.73 \\
        SAE steering (32 layer)                                     & 30.68 & 2.38 & 4.87 \\
        SAE steering (26+32 layers)                                     & \underline{19.88} & 3.70 & 7.58 \\
        \bottomrule
    \end{tabular}%
    }
\end{table}

\section{Conclusion}

In this paper, we investigated Whisper's internal representations for hallucination detection and mitigation. We showed that both raw activations and SAE latent representations encode linearly separable hallucination-related information, with classification performance improving toward deeper encoder layers. As few as 10--25 SAE latent dimensions suffice for near-optimal classification, confirming that hallucination-relevant information is concentrated in a sparse feature subset.

We proposed two steering strategies: activation-space steering and SAE latent-space additive steering. SAE-based steering consistently outperforms activation steering in hallucination reduction across all evaluated non-speech datasets, achieving HR of 19.88\% on UrbanSound8K, 27.05\% on WHAM!, and 33.92\% on FSD50k for Whisper large-v3 without any model fine-tuning, approaching the performance of Calm-Whisper fine-tuned (15.51\% on UrbanSound8K). For Whisper small, SAE steering reduces HR from 67.09\% to 8.68\% on UrbanSound8K, from 66.93\% to 4.68\% on WHAM!, and from 81.95\% to 26.12\% on FSD50k. Activation steering additionally improves WER on English speech for Whisper small, while CER on Chinese degrades under both strategies, partly due to the out-of-domain nature of Chinese speech for the SAE.

These results establish SAE-based steering as a viable fine-tuning-free approach to hallucination mitigation in ASR, with future work directed toward multilingual SAE training and multi-layer steering strategies.

\section{Acknowledgments}
The work of Vadim Popov and Tasnima Sadekova was prepared within the framework of the research project HSE-BR-2025-019 implemented as part of the Basic Research Program at HSE University.

\bibliographystyle{IEEEtran}
\bibliography{mybib}

\end{document}